\documentclass[12pt]{iopart}
\usepackage{iopams}
\usepackage[dvips]{color}
\def\bD{{\bi D}}
\def\bG{{\bi G}}
\def\bL{{\bi L}}
\def\bS{{\bi S}}
\def\bT{{\bi T}}
\def\bZ{{\bi Z}}
\def\cA{{\cal A}}
\def\cC{{\cal C}}
\def\cH{{\cal H}}
\def\cR{{\cal R}}
\def\cS{{\cal S}}
\def\cZ{{\cal Z}}
\def\ri{\mathrm{i}}
\def\gb{\beta}
\def\gd{\delta}

\def\gG{\Gamma}
\def\mutm{{\mu t_\mu m}}
\def\mut{{\mu t_\mu }}
\def\ket#1{\mid~\!{#1}~\!\rangle}
\def\bra#1{\langle~\!{#1}~\!\mid}
\def\modul#1{\vert~#1~\vert}
\def\d={\buildrel \rm def \over =}
\def\emline#1#2#3#4#5#6{%
       \put(#1,#2){\special{em:moveto}}%
       \put(#4,#5){\special{em:lineto}}}
\def\newpic#1{}

\begin{document}\jl{1}
\title{MODIFIED GROUP PROJECTORS: TIGHT BINDING METHOD}
\author{M Damnjanovi\'c\footnote[1]{E-mail:
yqoq@afrodita.rcub.bg.ac.yu}, T Vukovi\'c and I Milo\v sevi\'c}
\address{Faculty of Physics, University of Belgrade, P.O.Box 368,
 11001 Belgrade, Yugoslavia, http://www.ff.bg.ac.yu/qmf/qsg\_e.htm}

 \date{15 March 2000}
\begin{abstract}
Modified group projector technique for induced representations is a
powerful tool for calculation and symmetry quantum numbers
assignation of a tight binding Hamiltonian energy bands of crystals.
Namely, the induced type structure of such a Hamiltonian enables
efficient application of the procedure: only the interior
representations of the orbit stabilizers are to be considered. Then
the generalized Bloch eigen functions are obtained naturally by the
expansion to the whole state space. The method is applied to the
electronic $\pi$-bands of the single wall nanotubes: together with
dispersion relations, their complete symmetry assignation by the full
symmetry (line) groups and the corresponding symmetry-adapted eigen
function are found.
\end{abstract}
\submitted
\section{Introduction}\label{INTRO}
Let $S$ be the system with the symmetry group $\bG$. The group action
divides the system into (disjoint) orbits and the total system
state space decomposes onto the orbital subspaces. Each orbital space
is induced from the so called interior space of a single atom of that
orbit. Accordingly, in the state space $\cS=\cS_D$ the group acts by
the representation $D(\bG)$ decomposing onto the induced type orbital
subrepresentations. Typical examples are crystals; consisted of the
orbits of the translational group, their symmetry-adapted eigen
states are suitable linear combinations of the Bloch states,
corresponding to the energy band structure. The symmetry-adapted or
standard basis (SAB) can be found by the group projector technique.
Here we develop and apply its efficient modification \cite{YI-GP}
free of the difficulties emerging from the non-compactness of the
involved groups.

Let $D(\bG)$ decomposes onto the irreducible components
$D^{(\mu)}(\bG)$ as $D(\bG)=\oplus_\mu a_\mu D^{(\mu )}(\bG)$ ($a_\mu
=1,2,\dots$). The basis $\{\ket{\mutm}|\mu;\ t_\mu =1,\dots,a_\mu;\
m=1,\dots,|\mu|\}$ ($|\mu|$ is the dimension of $D^{(\mu )}(\bG)$)
of  $\cS_{D}$ is SAB if its vectors satisfy:
\begin{equation}\label{SAB}
D(g)\ket{\mutm}=\sum_{m'=1}^{|\mu |}D^{(\mu )}_{m'm}(g)\ket{\mutm'}.
\end{equation}
The direct product of two vectors from different $\bG$-spaces is a
fixed point of $\bG$ iff the transformations of the factors
mutually cancel, i.e. iff the factors transform oppositely. This
intuitive fact underlies the SAB construction by the modified group
projector technique. For each irreducible component $D^{(\mu)}(\bG)$
with $a_\mu>0$, the initial space $\cS_D$ is directly multiplied by
the dual $\cH^{(\mu)^*}$ of the space of $D^{(\mu)}(\bG)$. This space
carries the auxiliary representation $\gG^\mu(\bG)\d=D(\bG)\otimes
D^{(\mu)*}(\bG)$; its fixed point space, i.e. the range
$\cR(\bG(\gG^\mu))$ of the modified projector
$\bG(\gG^\mu)\d=\frac{1}{\modul{\bG}}\sum_g\gG^\mu(g)$, contains
exactly all the vectors with $\cS_D$ factors transforming according
to $D^{(\mu)}(\bG)$. Any basis $\{\ket{\mut}~|~\mu=1,\dots,a_\mu\}$
of $\cR(\bG(\gG^\mu))$ yields the $\mu$-part
of SAB as the partial scalar
products with the standard basis $\{\ket{\mu^*
m}~|~m=1,\dots,\modul{\mu}\}$ of $\cH^{(\mu^*)}$:
\begin{equation}\label{SABM}
\ket{\mutm}=\langle~\!{\mu^*m}~\!\mid~\!{\mut}~\!\rangle\quad
t_\mu=1,\dots,a_\mu;\ m=1,\dots,\modul{\mu}.
\end{equation}
If the basis is required to be simultaneously symmetry-adapted and
eigen basis (SAEB) of the Hamiltonian $H$ (naturally, $H$ commutes
with $D(\bG)$), the vectors $\ket{\mut}$ should be chosen as the
eigen vectors of $H\otimes I_\mu$ (the identity operator is denoted
by $I$). The method is especially powerful when $D(\bG)$ has
inductive structure $D(\bG)=\gd(\bS\uparrow \bG)\otimes d(\bG)$,
where $\gd(\bS)$ is representation (called interior) of the subgroup
$\bS$ and $d(\bG)$ any representation (called exterior) of $\bG$.
Eventual weak-direct product structure of $\bG$ enables further
simplifications.

In \sref{Stight} the SAB construction for multi orbit systems is
considered; it is shown how the interrelation of the tight-binding
Hamiltonian and the inductive structure of the state space
representation of the symmetry group yields an extremely simple way
to get the eigen energies together with the corresponding eigen
states. The results are applied to the single wall carbon nanotubes
(SWCT) \cite{IIJIMA,ELBi1} and their symmetry groups \cite{YITR} in
the \sref{Sswct}, to accomplish the full symmetry assignation of the
electronic bands and get the corresponding Bloch type
symmetry-adapted eigen functions.

\section{Tight binding Hamiltonian in the induced
state space}\label{Stight}
To introduce notation we briefly review the necessary results on the
modified group projector technique for induced representations
\cite{YI-GP}. Let $\bS$ be a subgroup in $\bG$ with the left
transversal $\bZ=\{z_p~|~p=0,\dots ,|\bZ|-1\}$ (by convention, $z_0$
is the identity element $e$ and $|\bZ|=\modul{\bG}/\modul{\bS}$).
Then for any fixed $g\in\bG$ and $z_p\in\bZ$ there are unique
$s(g,z_p)\in\bS$ and the index $p(g)$ satisfying
$g=z_ps(g,z_p)z^{-1}_{p(g)}$ (e.g. for $g=z_p$,
$s(z_p,z_p)=z_{p(z_p)}=e$, i.e. $p(z_p)=0$). Given the interior
representation $\gd(\bS)$ (in the space $\cS_{\gd}$ with the basis
$\{\ket{\psi}~|~\psi=1,\dots,\modul{\gd}\}$) and the exterior
representation $d(\bG)$ (in the space $\cH_d$), we look for SAB of
the induced type representation $D(\bG)=\gd(\bS\uparrow\bG)\otimes
d(\bG)$. The modified procedure deals with two auxiliary
representations for each irreducible component $D^{(\mu)}(\bG)$:
$\gG^\mu(\bG)=D(\bG)\otimes D^{(\mu)^*}(\bG)$ in the space
$\cS_{\gG^\mu}=\cS_D\otimes\cH^{(\mu^*)}=\oplus_{p}\cS_{p\gamma^\mu}$
(its operators are
$\gG^\mu(g)=\sum_pE^p_{p(g)}\otimes\gb^\mu_p\gamma^\mu(s(g,z_p))
\gb^{\mu^\dagger}_{p(g)}$), and its pulled down subgroup
representation $\gamma^\mu(\bS)=\gamma(\bS)\otimes
D^{(\mu)^*}(\bG\downarrow\bS)$ (here $\gamma(\bS)=\gd(\bS)\otimes
d(\bG\downarrow\bS)$; restricted representation denoted by
$\downarrow$) in
$\cS_{\gamma^\mu}=\cS_{\gd}\otimes\cH_d\otimes\cH^{(\mu)^*}$. The
spaces $\cS_{p\gamma^\mu}$ are the replicas of
$\cS_{0\gamma^\mu}=\cS_{\gamma^\mu}$. The matrices $E^p_q$ (with
vanishing all but $pq$-th element, which equals 1) are used to switch
between these spaces; together with the operators
$\gb^\mu_p=I_{\gd}\otimes d(z_p)\otimes D^{(\mu)^*}(z_p)$ in
$\cS_{\gamma^\mu}$ they give the transfer operators
$E^p_0\otimes\gb^\mu_p:\cS_{\gamma^\mu}\rightarrow\cS_{p\gamma^\mu}$.
The modified group projector $\bG(\gG^\mu)$ is essentially equivalent
to the pulled down projector $\bS(\gamma^\mu)$ (onto the fixed point
space of $\gamma^\mu(\bS)$) by the partial isometry
$B^\mu:\cS_{\gamma^\mu}\rightarrow\cS_{\gG^\mu}$:
\begin{equation}\label{BASIC}
\bG(\gG^\mu)=B^\mu\left\{E^0_0\otimes\bS(\gamma^\mu)\right\}B^{\mu^\dagger},
\quad B^\mu=\frac{1}{\sqrt{|Z|}}\sum_tE^t_0\otimes\gb^\mu_t.
\end{equation}
It has been shown that the basis $\ket{\mut}^0$ of the range of
$\bS(\gamma^\mu)$ determines the vectors $\ket{\mut}=
B^\mu\ket{\mut}^0$, resulting in the SAB $\ket{\mutm}$ given by
\eref{SABM}:
\begin{equation}\label{Estb}
\ket{\mutm}=\bra{\mu^*m}(B^\mu\ket{\mut}^0).
\end{equation}

We generalize these results to the case when
$D(\bG)=\oplus_PD^P(\bG)$ is direct sum of the several induced-type
representations from the subgroups $\bS^P$ (with the transversals
$\bZ^P$) and the interior representations $\gd^P(\bS^i)$, i.e.
$D^P(\bG)=\gd^P(\bS^P\uparrow\bG)\otimes d(\bG))$. It is
straightforward to show that the auxiliary representation
of the modified technique is already partially
reduced in the inductive spaces $\cS_{\gG^{P\mu}}$:
$\gG^\mu(\bG)=D(\bG)\otimes D^{(\mu)^*}(\bG)=
\oplus_P\gG^{P\mu}(\bG)$, where $\gG^{P\mu}(\bG)=D^P(\bG)\otimes
D^{(\mu)^*}(\bG)$. Therefore it is pulled down
into the space $\cS_{\gamma^\mu}=\oplus_P\cS_{\gamma^{P\mu}}$ for each
component $\gG^{P\mu}$ independently, and the transfer operator,
mapping $\cS_{\gamma^\mu}$ into $\cS_{\gG^\mu}$ is obvious
generalization of \eref{BASIC}:
\begin{equation}\label{EBtrans}
B^\mu=\sum_PB^{P\mu},\quad B^{P\mu}=
\frac{1}{\sqrt{|\bZ^P|}}\sum_{p\in\bZ^P}E^{Pp}_{P0}\otimes\gb^{P\mu}_{p},
\end{equation}
with $\gb^{P\mu}_{p}=I_{\gd^P}\otimes d(z^P_{p})\otimes
D^{(\mu)^*}(z^P_{p})$. Finally, the pulled down projector is
found:
\begin{equation}\label{EmultiBASIC}
\bG^{\downarrow}(\gG^\mu)=
\sum_PB^{P\mu^\dagger}\bG(\gG^{P\mu})B^{P\mu}=
\sum_PE^{P0}_{P0}\otimes\bS^P(\gamma^{P\mu})
\end{equation}
Note the possibility to obtain the standard basis independently in
each subspace $\cS_{D^P}$ using \eref{Estb}.

Further, let the Hamiltonian be given and SAEB is looked for.
Preliminary, we show that $H$ can be always chosen to act trivially
in $\cH_d$. Suppose $d(\bG)=d'(\bG)\otimes d_0(\bG)$, and $H$ acts
trivially only in $\cH_{d_0}$. Due to the equivalence of the
representations $(\gd(\bS)\uparrow\bG)\otimes d'(\bG)$ and
$(\gd(\bS)\otimes d'(\bG\downarrow\bS))\uparrow\bG$ the non-trivial
part can be absorbed in the interior representation, using
$\gd(\bS)\otimes d'(\bG\downarrow\bS)$ instead of $\gd(\bS)$. Thus,
the Hamiltonian has the form $H=H'\otimes I_d$ providing its
triviality in the external space $\cH_d$. Then, by the modified
prescription the auxiliary operator $H_\mu=H\otimes I_\mu$ (in
$\cS_D\otimes\cH^{{(\mu)}^*}$) is pulled down to the space
$\cS_{\gamma^\mu}$ by $B^\mu$ giving
\begin{equation}\label{EHdown}
H^{\downarrow}_\mu\d= B^{\mu^\dagger}H_\mu B^\mu=
\sum_{PQ}E^{P0}_{Q0}\otimes \frac{\sum_{pq}\gb^{P\mu^\dagger}_{p}
h^{Pp}_{Qq}\gb^{Q\mu}_{q}}{\sqrt{\modul{\bZ^P}\modul{\bZ^Q}}},
\end{equation}
where
\begin{equation}\label{EHmat}
h^{Pp}_{Qq}=(\sum_{\phi,\psi}h^{Pp\phi}_{Qq\psi}\ket{\phi}\bra{\psi})
\otimes I_d\otimes I_\mu.
\end{equation}
are the (rectangular) submatrices in the decomposition
$H_\mu=\sum_{PQ}\sum_{pq}E^{Pp}_{Qq}\otimes h^{Pp}_{Qq}$.
Commutativity of $H$ with $D(\bG)$ interrelates these matrices: for
the transversal elements the conditions $[H,D(z^P_p)]=0$ give
$h^{Pp}_{Qq}=(\gd^P(s^P(z^Q_{q},z^P_p))\otimes I_d\otimes I_\mu
)h^{Pp(z^Q_{q})}_{Q0}$, and \eref{EHdown} becomes
\begin{equation}\label{EHdowngen}
H^{\downarrow}_\mu=\sum_{PQ}E^{P0}_{Q0}\otimes\frac{\sum_{pq}
\gamma^{P\mu^\dagger}(s^P(z^{Q^{-1}}_q,z^P_p))\gb^{P\mu^\dagger}_{p}
h^{Pp}_{Q0}}{\sqrt{\modul{\bZ^P}\modul{\bZ^Q}}}.
\end{equation}
$H^\downarrow_\mu$ commutes with the projector
$\bG^{\downarrow}(\gG^\mu)$, and the vectors $\ket{\mut}^0$ should be
chosen as the eigen vectors of $H^\downarrow_\mu$ from the range of
$\bS(\gamma^\mu)$:
$H^{\downarrow}_\mu\ket{\mut}^0=\epsilon_{\mut}\ket{\mut}^0$ and
$\bG^\downarrow(\gG^\mu)\ket{\mut}^0=\ket{\mut}^0$. Finally, the
expansion \eref{Estb} gives the vectors $\ket{\mutm}$. It can be
proved that the obtained basis is the wanted SAEB of $H$, with the
same eigen values $\epsilon_{\mut}$. To differ from the general
standard basis, it may turn out that the vectors $\ket{\mut}^0$ are
inevitably combinations of the vectors from various
$\cS_{\gamma^{P\mu}}$. Then the SAEB mixes the states of different
spaces $\cS_{D^P}$, too.

Especially, if all the subgroups $\bS^P$ have common transversal
$\bZ$, and $\bZ$ is itself a subgroup of $\bG$ (i.e. when
$\bG=\bZ\bS^P$ --- the weak direct product), then $s^P(z^{-1}_q,z_p)=e$
for any $p$ and $q$, simplifying the pulled down Hamiltonian:
\begin{equation}\label{EHdownsubg}
H^\downarrow_\mu=
\sum_{PQ}E^{P0}_{Q0}\otimes\sum_{p}\gb^{\mu^\dagger}_ph^{Pp}_{Q0}.
\end{equation}

Let the system $S$ (e.g. molecule, crystal, multi layer, polymer etc.)
with the symmetry group $\bG$ consists of $\bG$ orbits $S^P$
($P=1,2,\dots$). The corresponding stabilizers and transversals are
$\bS^P$ and $\bZ^P$. Obviously, choosing an atom on the orbit for the
initial one, the transversal elements $z^P_p$ biuniquely correspond
to the atoms of that orbit. Thus the atoms of the system are
enumerated by the pair $(Pp)$ of the orbit and the transversal
element indices. Each orbit defines the representation $\gd^P(\bS^P)$
related to the studied property. For example, when the vibrational
modes and the spin-waves are studied, to each atom the polar- and the
axial-vector representation are associated.

In the tight-binding approximation each atom of the
orbit $S^P$ contributes to the relevant state space (of
valent electrons) $\cS$ by $\gd^P$ atomic orbitals
$\ket{(Pp)\psi}$ ($\psi=1,\dots,\gd^P$) spanning the interior space
$\cS_{p\gd^P}$; the action of the stabilizer element $s^P$ on
$\ket{(Pp)\psi}$ gives the linear combination of the orbitals from
the same atom,
$\sum_{\phi=1}^{\gd^P}\gd^P_{\phi\psi}(s^P)\ket{(Pp)\phi}$,
defining the interior representation $\gd^P(\bS^P)$ (thus
$\gd^P=\modul{\gd^P(\bS^P)}$). In this problem there is no external
representation $d(\bG)$, i.e. in the previous expressions it should
be omitted (substituted by the trivial representation). Finally, the
direct sum of the interior spaces gives the total state space
$\cS=\cS_D$ where the symmetry group acts by the inductive
representation $D(\bG)=\oplus_P\gd^P(\bS^P\uparrow\bG)$.

Within this model, each atom $(Pp)$
interacts with $N^P$ neighbors $(Pp;n)$ ($n=1,\dots,N^P$) by a
coupling coefficients $V^{Pn}_{\psi\phi}$ between the valent orbitals
(the number of the interacting neighbors and
the interactions with them is same for all the atoms of the same
orbit due to symmetry). So, the Hamiltonian is
\begin{equation}\label{EtbaH}
H=\sum_{P}\sum_{n=1}^{N^P}\sum_{p=0}^{|\bZ^P|-1}\sum_{\psi=1}^{\gd^P}
\sum_{\phi=1}^{\gd^{[Pp;n]}}V^{Pn}_{\phi\psi}
\ket{(Pp;n)\phi}\bra{(Pp)\psi},
\end{equation}
where $[Pp;n]$ is the orbit of the neighbor $(Pp;n)$. Therefore,
using in \eref{EHmat} the matrix elements
$h^{Pp\phi}_{Q0\psi}=\sum_{n=1}^{N^Q}V^{Qn}_{\phi\psi}\gd^{(Pp)}_{(Q0;n)}$
($\gd^i_j$ is the Kronecker delta) determined by \eref{EtbaH}, the
pulled down Hamiltonian becomes
\begin{equation}\label{Etbadown}
\fl H^\downarrow_\mu=\sum_Q\sum_{n=1}^{N^Q}E^{[Q0;n]0}_{Q0}\otimes
\left\{\gamma^{[Q0;n]\mu}_{Q(Q0;n)}\gb^{[Q0;n]\mu^\dagger}_{(Q0;n)}
(\sum_{\phi=1}^{\gd^{[Q0;n]}}\sum_{\psi=1}^{\gd^Q}
V^{Qn}_{\phi\psi}\ket{\phi}\bra{\psi}\otimes I_\mu)\right\},
\end{equation}
where
$\gamma^{P\mu}_{Qp}=\sum_q\gamma^{P\mu^\dagger}(s^{P}(z^{Q^{-1}}_q,z^{P}_{p}))/
\sqrt{\modul{\bZ^Q}\modul{\bZ^{[Q0;n]}}}$.
Finally, when there is common transversal being a
subgroup in $\bG$, \eref{EHdownsubg} reduces to the simple sum over
neighbors:
\begin{equation}\label{Etbadownsubg}
H^\downarrow_\mu=\sum_Q\sum_{n=1}^{N^Q}E^{[Q0;n]0}_{Q0}\otimes
\left\{
\gb^{[Q0;n]\mu^\dagger}_{(Q0;n)}(
\sum_{\phi=1}^{\gd^{[Q0;n]}}\sum_{\psi=1}^{\gd^Q}
V^{Qn}_{\phi\psi}\ket{\phi}\bra{\psi}\otimes I_\mu)\right\}.
\end{equation}

\section{Single-wall carbon nanotube bands}\label{Sswct}
Symmetry groups $\bL_\cC$ of chiral $(n_1,n_2)$,
and $\bL_{\cZ\cA}$ of zig-zag $(n,0)$
and armchair $(n,n)$ SWCT ($\cC$, $\cZ$ and $\cA$ tubes for short)
are the line groups  \cite{YITR} given  in the factorized and
international notation
\begin{equation}\label{Elinegroups}
\bL_\cC=\bT^r_q\bD_n=\bL q_p22,\quad
\bL_{\cZ\cA}=\bT^1_{2n}\bD_{nh}=\bL2n_n/mcm,
\end{equation}
with the translational period $a=a_0\sqrt{3q/2n{\cal R}}$. Here
$a_0=2.461${\AA} is the honeycomb lattice period, while $n$, $q$,
$p$, $r$ and $\cR$ are defined in terms of $n_1$ and $n_2$. The
parameters $\tilde{q}$ and $q$ are even since
$\tilde{q}=q/n=2\pmod{12}$, while $n$ and $p$ are even iff both $n_1$
and $n_2$ are even; also $q>10$ for the realistic tubes (having
diameters $D=a_0\sqrt{{\cal R} nq/2}/\pi$ greater than 3.4{\AA}).

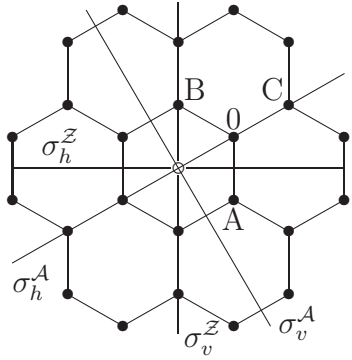
\begin{figure}\centering
\unitlength=0.70mm \special{em:linewidth 0.4pt} \linethickness{0.4pt}
\begin{picture}(68.50,65.00)
\put(4.50,28.00){\circle*{2.00}} \put(4.50,40.00){\circle*{2.00}}
\
\put(15.00,10.00){\circle*{2.00}} \put(15.00,22.00){\circle*{2.00}}
\put(15.00,46.00){\circle*{2.00}} \put(15.00,58.00){\circle*{2.00}}
\
\put(25.50,4.00){\circle*{2.00}} \put(25.50,28.00){\circle*{2.00}}
\put(25.50,40.00){\circle*{2.00}} \put(25.50,64.00){\circle*{2.00}}
\
\put(36.00,10.00){\circle*{2.00}} \put(36.00,22.00){\circle*{2.00}}
\put(36.00,46.00){\circle*{2.00}} \put(36.00,58.00){\circle*{2.00}}
\
\put(46.50,4.00){\circle*{2.00}} \put(46.50,28.00){\circle*{2.00}}
\put(46.50,40.00){\circle*{2.00}} \put(46.50,64.00){\circle*{2.00}}
\
\put(57.00,10.00){\circle*{2.00}} \put(57.00,22.00){\circle*{2.00}}
\put(57.00,46.00){\circle*{2.00}} \put(57.00,58.00){\circle*{2.00}}
\
\put(67.50,28.00){\circle*{2.00}} \put(67.50,40.00){\circle*{2.00}}
\
\emline{4.50}{40.00}{41}{15.00}{46.00}{42}
\
\emline{4.50}{28.00}{43}{15.00}{22.00}{44}
\
\emline{15.00}{58.00}{3}{25.50}{64.00}{4}
\
\emline{15.00}{10.00}{5}{25.50}{4.00}{6}
\emline{15.00}{46.00}{7}{25.50}{40.00}{8}
\
\emline{25.50}{4.00}{9}{36.00}{10.00}{10}
\emline{25.50}{40.00}{11}{36.00}{46.00}{12}
\
\emline{25.50}{28.00}{13}{36.00}{22.00}{14}
\emline{25.50}{64.00}{15}{36.00}{58.00}{16}
\
\emline{36.00}{22.00}{17}{46.50}{28.00}{18}
\emline{36.00}{58.00}{19}{46.50}{64.00}{20}
\
\emline{36.00}{10.00}{21}{46.50}{4.00}{22}
\emline{36.00}{46.00}{23}{46.50}{40.00}{24}
\
\emline{46.50}{4.00}{25}{57.00}{10.00}{26}
\
\emline{46.50}{28.00}{29}{57.00}{22.00}{30}
\emline{46.50}{64.00}{31}{57.00}{58.00}{32}
\
\emline{57.00}{22.00}{33}{67.50}{28.00}{34}
\
\emline{57.00}{46.00}{35}{67.50}{40.00}{36}
\
\put(4.50,28.00){\line(0,1){12.00}}
\
\put(15.00,10.00){\line(0,1){12.00}}
\put(15.00,46.00){\line(0,1){12.00}}
\
\put(25.50,28.00){\line(0,1){12.00}}
\
\put(36.00,10.00){\line(0,1){12.00}}
\put(36.00,46.00){\line(0,1){12.00}}
\
\put(46.50,28.00){\line(0,1){12.00}}
\
\put(57.00,10.00){\line(0,1){12.00}}
\put(57.00,46.00){\line(0,1){12.00}}
\
\put(67.50,28.00){\line(0,1){12.00}}
\
 \put(36.00,34.00){\circle{2.00}}
 \put(4.50,34.00){\line(1,0){30}}
 \put(37.50,34.00){\line(1,0){30}}
 \put(36.0,35.50){\line(0,1){30}}
 \put(36.0,32.50){\line(0,-1){30}}
 \emline{4.50}{16}{1}{67.5}{52}{2}
 \emline{18.50}{64}{1}{53.5}{4}{2}
 \put(10,35.0){\makebox(0,0)[lb]{$\sigma^\cZ_h$}}
 \put(37,5.0){\makebox(0,0)[lt]{$\sigma^\cZ_v$}}
 \put(55,4.5){\makebox(0,0)[lc]{$\sigma^\cA_v$}}
 \put(4.5,15.5){\makebox(0,0)[lt]{$\sigma^\cA_h$}}
 \put(46.5,41.5){\makebox(0,0)[cb]{0}}\put(46.5,26.5){\makebox(0,0)[ct]{A}}
 \put(37,47){\makebox(0,0)[lb]{B}}\put(56,47){\makebox(0,0)[rb]{C}}
\end{picture}
\caption{Nearest neighbors of the C-atom atom denoted by 0, are the
atoms 1, 2 and 3. Perpendicular to the figure at $\circ$ is $U$-axis
being also $x$-axis of the coordinate system;
$\sigma^{\cZ/\cA}_{h/v}$ stands for vertical and horizontal mirror
planes of $\cZ$ and $\cA$ tubes.}\label{Fneigh3}
\end{figure}

From the factorized notation  it follows that these groups are
weak-direct products of the cyclic groups generated by
$(C^{r}_q|a/\tilde{q})$, $C_n$ and $U$ for $\bL_\cC$, and
additionally $\sigma_x$ for $\bL_{\cZ\cA}$ (see \fref{Fneigh3}). Thus, their
elements are monomials
\begin{equation}\label{Eelems}
\ell(t,s,u,v)=(C^{r}_q|\frac{a}{\tilde{q}})^tC^s_nU^u\sigma^v_x
\end{equation}
for $t=0,\pm1,\dots$, $s=0,\dots,n-1$, $u=0,1$, and
$v=0$ in $\bL_\cC$ and $v=0,1$ in $\bL_{\cZ\cA}$.

Each SWCT is a single orbit of its line group. The type of the orbit
according to the classification of \cite{IY}, its stabilizer and the
transversal are given in the \tref{Torbits}. In all the cases the
transversal is the group $\bT^r_q\bD_n$ used to enumerate the atoms:
initial atom $\mbox{C}_{000}$ has the radius vector in the
cylindrical coordinate system of \fref{Fneigh3}:
\begin{equation}\label{ECinit}
{\bi r}_{000}=(\frac{D}{2},\phi_0,z_0),\quad
\phi_0=2\pi\frac{n_1+n_2}{nq\cR},\quad
 z_0=\frac{n_1-n_2}{\sqrt{6nq\cR}}a_0.
\end{equation}
The transversal element $\ell(t,s,u,0)$ maps C$_{000}$ to the atom
C$_{tsu}$ with the coordinates
${\bi r}_{tsu}=(D/2,(-1)^u\phi_0+2\pi(t/q+s/n),(-1)^uz_0+tna/q)$.
Experimental data verified the validity of the simple tight-binding
nearest neighbors model. Each C atom contributes by one $p^\bot$
(i.e. graphene $p_z$) orbital $\ket{(tsu)}$ (the notation is
simplified: the capital and the Greek indices for orbits and interior
space vectors, taking on only the value 1, are omitted). The orbital
$\ket{(tsu)}$ spans the (one dimensional) interior space carrying the
trivial stabilizer representation $\gd(\bS)=1(\bS)$ (triviality is
obvious for $\cC$ tubes, and for the $\cZ$ and $\cA$ ones it follows
from the polar-vector type of the $Y^m_{l=1}$ spherical harmonics
involved in $p^\bot$ orbital). Thus in the total independent electron
space $\cS$, being spanned by all these orbitals, the symmetry group
acts by the induced representation $D(\bG)=1(\bS\uparrow\bG)$. The
pulled down modified projector with the irreducible representation
$D^{(\mu)}(\bG)$ (see appendix) becomes
$\bS(D^{(\mu)^*})=\sum_{s\in\bS}D^{(\mu)^*}(s)/\modul{\bS}$, and the
dimension of its range gives the frequency number of $D^{(\mu)}(\bG)$
in $D(\bG)$.
\begin{table}\centering
\caption{Orbits and neighbors of SWCT. For $\cC$, $\cZ$ and $\cA$
tubes their line group, orbit type, stabilizer and transversal group
are in the columns $\bG$, $O$, $\bS$ and $\bZ$. Then follow the
parameters $t_n$ and $s_n$ ($u_n=1$) expressing the three nearest
neighbors of the initial atom in the form
$(000;n)=\mbox{C}_{t_ns_n1}$.} \label{Torbits}
\footnotesize\rm
\begin{tabular}{@{}llllllll}\br
&$\bG$&O&$\bS$&$\bZ$&$(000;1)$&$(000;2)$&$(000;3)$\\ \mr
$\cC$&$\bL_\cC$&$a_1$&$\{e\}$&$\bT^r_q\bD_n$&$
{t_1=-\frac{n_2}{n}\atop s_1=\frac{2n_1+(1+r\cR)n_2}{q\cR}}$&$
{t_2= \frac{n_1}{n}\atop s_2=\frac{(1-r\cR)n_1+2n_2}{q\cR}}$&$
{t_3=t_1+t_2       \atop s_3=s_1+s_2}$\\
$\cZ$&$\bL_{\cZ\cA}$&$b_1$&$\{e,C_n\sigma_x\}$&$\bT^1_{2n}\bD_n$&$
t_1=0\ s_1=1$&$t_2=1\ s_2=0$&$t_3=1\ s_3=1$\\
$\cA$&$\bL_{\cZ\cA}$&$d_1$&$\{e,\sigma_h\}$&$\bT^1_{2n}\bD_n$&$
t_1=-1\ s_1=1$&$t_2=1\ s_2=0$&$t_3=0\ s_3=1$\\ \br
\end{tabular}
\end{table}
The Hamiltonian confines the interaction to the 3 nearest
neighbors (see \fref{Fneigh3}). Therefore, the pulled down
Hamiltonian reads:
\begin{equation}\label{EtbadownSWCT}
H^\downarrow_\mu=\sum_{n=0}^{3}V_n
D^{(\mu)^{*\dagger}}(\ell(t_n,s_n,u_n,0)).
\end{equation}
Here, $\ell(t_n,s_n,u_n,0)$ is the transversal element which maps the
initial atom into its neighbor $(000;n)$; it is given in
\tref{Torbits}. By $n=0$ the initial atom itself is included (the
diagonal term of the Hamiltonian); nevertheless, its contribution to
the energy is reduced to the additive constant $V_0$, and hereafter
it is omitted. Further, since the rolling up induced distortions of
the honeycomb lattice are nearly homogeneous, all the remaining
coupling constants are approximately equal: $V_1=V_2=V_3=V$
(estimated between $-2.7$eV and $-2.5$eV).

\begin{table}
\caption{Bands and symmetry-adapted eigen vectors of the carbon
nanotubes. For each irreducible component $D^{(\mu)}(\bL)$ of
$D(\bL)=1(\bS\uparrow\bL)$ its frequency number $a_\mu$, energies
$\epsilon_{\mut}$,the transfer operator $\gb^\mu_{tsu}$ and the
pulled down standard eigen basis $\ket{\mut}^0$ are given. For the
matrices $M_2,K_2,O_2,M_4, K_4$ and $O_4$ see the appendix.}
\label{Tbands}\begin{indented}\item[]\begin{tabular}{@{}llllll}\br
$\cC$&$a_\mu$&$\epsilon_{\mut}/V$&$\gb^\mu_{tsu}$&$\ket{\mut}^0$\\ \mr
${_0}A^\Pi_m$&$1$&$\Pi(1+2\e^{\ri\frac{m\pi}{q}})$&$\Pi^u \e^{\ri
m(rt+s\tilde{q})\frac{2\pi}{q}}$&$\ket{0m\Pi}$\\
${_\pi}A^\Pi_m$&$1$&$-\Pi$&$\Pi^u\e^{\ri(m(rt+s\tilde{q})\frac{2\pi}{q}
+\frac{t\pi}{\tilde{q}})}$&$\ket{\pi m\Pi}$\\
${_k}E_m$&$2$&\eref{EnbandsgenC}&$
K_2(\frac{kt}{\tilde{q}})M_2(m\frac{tr+s\tilde{q}}{\tilde{q}})O^u_2$&$
\ket{km},\ket{-k,-m}$\\ \mr
${_0}A^\Pi_{\tilde{m}}$&$1$&$\Pi(1+2\e^{\ri\frac{m\pi}{n}})$&$\Pi^u
\e^{\ri\tilde{m}s\frac{2\pi}{n}}$&$\ket{0\tilde{m}\Pi}$\\
${_{\tilde{\pi}}}A^\Pi_{\tilde{m}}$&$1$&$-\Pi$&$\Pi^u(-1)^t
\e^{\ri\tilde{m}s\frac{2\pi}{n}}$&$\ket{\tilde{\pi}\tilde{m}\Pi}$\\
${_{\tilde{k}}}E_{\tilde{m}}$&$2$&\eref{EnbandsgenC}&$
K_2(\frac{\tilde{k}}{\tilde{q}})M_2(s\tilde{m})O^u_2$&$
\ket{\tilde{k}{m}},\ket{-\tilde{k},-\tilde{m}}$\\ \br
\end{tabular}\item[]\begin{tabular}{@{}llllll}\br
$\cZ$&$a_\mu$&$\epsilon_{\mut}/V$&$\gb^\mu_{tsu}$&$\ket{\mut}^0$\\ \mr
${_0}A^\Pi_m$&1&$\Pi(1+2\e^{\ri\frac{m\pi}{n}})$&$
    \Pi^u\e^{\ri\frac{m\pi}{n}t}$&$\ket{0m\Pi A}$\\
${_0}E^\Pi_m$&$1$&$\Pi(1+2\cos\frac{m\pi}{n})$&$
\Pi^uM_2(-\frac{2s+t}{2}m)O^u_2$&$
\frac{\ket{0m\Pi}+\e^{\ri\frac{m2\pi}{n}}\ket{0,-m,\Pi}}{\sqrt{2}}$\\
${_k}E^A_m$&$2$&$\pm\sqrt{5+4\e^{\ri\frac{m\pi}{n}}\cos\frac{ka}{2}}$&$
\e^{\ri\frac{m\pi}{n}t}K_2(\frac{kt}{2})O^u_2$&$
\ket{kmA},\ket{-k,m,A}$\\
${_\pi}E^\Pi_\frac{n}{2}$&$1$&$-\Pi$&$
\ri^t(-1)^s\Pi^uK_2(\frac{t\pi}{2a})$&$\frac{
\ket{\frac{\pi}{a},\frac{n}{2},\Pi}-
\ket{\frac{\pi}{a},-\frac{n}{2},\Pi}}{\sqrt{2}}$\\
${_k}G_m$&$2$&\eref{EnbandsgenZ}&$K_4(\frac{kt}{2},m\frac{t+2s}{2})O^u_4$&$
{\frac{\ket{km}+\e^{-\ri\frac{m2\pi}{n}}\ket{k,-m}}{\sqrt{2}}\atop
\frac{\ket{-k,m}+\e^{-\ri\frac{m2\pi}{n}}\ket{-k,-m}}{\sqrt{2}}}$\\
\br
\end{tabular}\item[]\begin{tabular}{@{}llllll}\br
$\cA$&$a_\mu$&$\epsilon_{\mut}/V$&$\gb^\mu_{tsu}$&$\ket{\mut}^0$\\ \mr
${_0}\Pi^+_m$&1&$1+2\e^{\ri\frac{m\pi}{n}}$&$
    \Pi^u\e^{\ri\frac{m\pi}{n}t}$&$\ket{0,m,+,\Pi}$\\
${_0}E^+_m$&$2$&$\pm\sqrt{5+4\cos\frac{m\pi}{n}}$&$
\Pi^uM_2(-\frac{2s+t}{2}m)O^u_2$&$
\ket{0m+},\ket{0,-m,+}$\\
${_k}E^\Pi_m$&$1$&$\Pi(1+2\e^{\ri\frac{m\pi}{n}}\cos\frac{ka}{2})$&$
\Pi^u\e^{\ri\frac{m\pi}{n}t}K_2(\frac{kt}{2})O^u_2$&$
\frac{\ket{km\Pi}+\ket{-k,m,\Pi}}{\sqrt{2}}$\\
${_\pi}E^\Pi_\frac{n}{2}$&$1$&$-3\Pi$&$
\ri^t(-1)^s(\pm1)^uK_2(\frac{t\pi}{2a})$&$\frac{
\ket{\frac{\pi}{a},\frac{n}{2},\Pi}+\Pi
\ket{\frac{\pi}{a},-\frac{n}{2},\Pi}}{\sqrt{2}}$\\
${_k}G_m$&$2$&\eref{EnbandsgenA}&$K_4(\frac{kt}{2},m\frac{t+2s}{2})O^u_4$&$
{\frac{\ket{km}+\ket{k,-m}}{\sqrt{2}}\atop
\frac{\ket{-k,m}+\ket{-k,-m}}{\sqrt{2}}}$\\ \br
\end{tabular}\end{indented}\end{table}
Finally, using the irreducible representations from the appendix and
the data from \tref{Torbits}, the Hamiltonian and the
modified projectors are pulled down and the results are given in
\tref{Tbands}. The most general dispersion relations for the bands of
$\cC$, $\cZ$ and $\cA$ tubes are respectively:
\numparts\label{Enbandsgen}\begin{eqnarray}
\epsilon^\pm_{E_m}(k)=\pm V\sqrt{1+8\cos\frac{\psi_A}{2}
\cos\frac{\psi_B}{2}\cos\frac{\psi_A-\psi_B}{2}},\label{EnbandsgenC}\\
\epsilon^\pm_{G_m}(k)=\pm V\sqrt{1+4\cos\frac{ka}{2}\cos\frac{m\pi}{n}
+4\cos^2\frac{m\pi}{n}},\label{EnbandsgenZ}\\
\epsilon^\pm_{G_m}(k)=\pm V\sqrt{1+4\cos\frac{ka}{2}\cos\frac{m\pi}{n}
+4\cos^2\frac{ka}{2}}\label{EnbandsgenA}.
\end{eqnarray}\endnumparts
The angles used in \eref{EnbandsgenC} are
\begin{equation}
\psi_A=-ka\frac{n_2}{q}+2\pi m\frac{2n_1+n_2}{qn\cR},\quad
\psi_B=ka\frac{n_1}{q}+2\pi m\frac{n_1+2n_2}{qn\cR}.
\end{equation}
Substituting them in \eref{EnbandsgenC} by
\begin{equation}
\fl\tilde{\psi}_A=-\tilde{k}a\frac{n_2}{q}+2\pi\tilde{m}
\frac{2n_1+(1+r\cR)n_2}{qn\cR},\quad
\tilde{\psi}_B=\tilde{k}a\frac{n_1}{q}+
2\pi\tilde{m}\frac{(1-r\cR)n_1+2n_2}{qn\cR},
\end{equation}
the $\tilde{k}\tilde{m}$ bands are obtained. To find the standard
basis $\ket{\mutm}$ it suffices to apply \eref{Estb}, with
$\gb^{\mu}_{tsu}$ and $\ket{\mut}^0$ given in \tref{Tbands}.
For example, the bands of the representations ${_k}E^{\Pi^v}_m$ of
the $\cZ$ and $\cA$ tubes corresponds to the symmetry
adapted generalized Bloch states:
\begin{equation}\label{EBloch}
\fl\ket{km\Pi^v}=\sum_{ts}\e^{\ri(m\frac{\pi}{n}+k\frac{a}{2})t}\ket{(ts0)},\
\ket{-k,m,\Pi^v}=\Pi^v\sum_{ts}\e^{\ri(m\frac{\pi}{n}-k\frac{a}{2})t}\ket{(ts1)}.
\end{equation}

\section{Summary}
The modified group projector technique is applied to find the
symmetry adapted basis for the sum $D(\bG)=\oplus_P
\gd^P(\bS^P\uparrow\bG)\otimes d(\bG)$ of the induced
interior representations $\gd^P(\bS^P)$ of arbitrary subgroups
$\bS^P$ of $\bG$ directly multiplied by the exterior representation
$d(\bG)$. It is shown that this basis is expanded basis
$\ket{\mut}^0$ of the range of the pulled down projector
$\bG^\downarrow(\gG^\mu)=\sum_PE^{P0}_{P0}\gamma^{P\mu}(\bS^P)$ in the
low dimensional auxiliary space (summed interior spaces
multiplied by the exterior and dual irreducible space). The same
procedure is applied when the standard basis, being also the eigen
basis of the Hamiltonian $H$ (acting trivially in the exterior space)
is looked for: then in the range of $\bG^\downarrow(\gG^\mu)$ the
eigen vectors $\ket{\mut}^0$ of the pulled down Hamiltonian
$H^\downarrow_\mu$ are to be found. In particular, for the general
tight-binding Hamiltonian the operators $H^\downarrow_\mu$ are found.

It is interesting to understand the physical meaning of the operator
$\gamma^{P\mu}_{Qp}$ appearing in \eref{Etbadown}. In fact,
the different transversals $\bZ^P$ and $\bZ^Q$ arrange the atoms on
the corresponding orbits differently. Even when the two
orbits have the same transversal, since in general it is not a
subgroup, the successive elements are not arranged homogeneously.
Thus, the sum over the orbit $S^Q$ involved in $\gamma^{P\mu}_{Qp}$ gives
the average relative positions between $S^Q$ and $S^P$.
The terms in the sum are mutually equal when the relative positions
of the atoms on the two orbits are constant, provided by the
homogeneous action of the same transversal when it is a subgroup.

These results enable simple calculation of the crystal energy
bands, with automatic assignation of the bands by the symmetry
quantum numbers. The method is applied to find the
energy bands and corresponding eigen functions of the carbon nanotubes
\cite{IIJIMA}. Although the bands dispersion relations have been
already calculated in the literature both for $km$
\cite{ELBi1,ELBi2}, and $\tilde{k}\tilde{m}$ quantum numbers
\cite{ELBy1,ELBy2,ELBy3,ELBy4}, only the helical part of the symmetry
group has been used, neglecting the parities. Such incomplete
band assignation can produce errors in studying
various processes, since the selection rules incorporating parities
are more severe, forbidding some otherwise allowed interband
transitions. Also the generalized Bloch eigen functions have the most
precise form only when the full symmetry is used. The necessity for
full symmetry group treatment is enforced by the fact that the peaks
of the density of states mostly correspond to the even or odd
states, i.e. to the representations with parities. This and some
other symmetry related questions (e.g. topology of bands,
including band sticking and metallic properties) will be analyzed
elsewhere.

\appendix
\section{
Irreducible representations of $\bL_\cC$ and $\bL_{\cZ\cA}$}
There are two physically based classification of the irreducible
representations of the nonsymorphic line groups (like the considered
ones), differing in used quasi momenta: quantum numbers labeling
representations may be either $k$ and $m$ of linear and angular quasi
momenta, or $\tilde{k}$ and $\tilde{m}$ of helical (includes linear
and a part of angular) and remaining angular momemnta. As for the
$\cC$ tubes both are used in literature (see e.g. \cite{ELBi1,ELBi2}
for $km$ and \cite{ELBy1,ELBy2,ELBy3,ELBy4} for $\tilde{k}\tilde{m}$
classification), while for the $\cZ$ and $\cA$ tubes
only the former one, as it has been done in this paper.
\begin{table}
\caption{Irreducible representations of $\bL_\cC$ classified by
the $km$ and $\tilde{k}\tilde{m}$ quantum numbers. For the
irreducible representations denoted in the first column, possible
values of $k$ and $m$ ($\tilde{k}$ and $\tilde{m}$) are given in the
second column, and then the matrices of the generators follow.
Finally, SAB is given in terms of quantum numbers. Only for
integer $m=n/2,q/2,-p/2,(q-p)/2$ the corresponding representations
appear. (*) $k=0$ with
$m\in(0,\frac{q}{2})$, and $k=\frac{\pi}{a}$ with
$m\in(-\frac{p}{2},\frac{q-p}{2})$ and $k\in(0,\frac{\pi}{a})$ with
$m\in(-\frac{q}{2},\frac{q}{2}]$.} \label{TCirs5IY}
\footnotesize\rm\begin{indented}\item[]
\begin{tabular}{@{}llllll}\br
IR&$(k,m)$&$(C^r_q|\frac{a}{\tilde{q}})$&$C_n$&$U$&SAB\\ \mr
${_k}A^{\Pi^U}_m$&${k=0,m=0,\frac{q}{2}\atop
               k=\frac{\pi}{a},m=-\frac{p}{2},\frac{q-p}{2}}$&$
\e^{\ri(mr\frac{2\pi}{q}+k\frac{a}{\tilde{q}})}$&$\e^{\ri
m\frac{2\pi}{n}}$&$\Pi^U$&$\ket{km\Pi^U}$\\
${_k}E_m$&(*)&$M_2(\frac{mr}{\tilde{q}})K_2(\frac{k}{\tilde{q}})$&$
M_2(m)$&$O_2$&${\ket{km}\atop\ket{-k,-m}}$\\ \hline
${_{\tilde{k}}}A^{\Pi^U}_{\tilde{m}}$&$
\tilde{k}=0,\tilde{q}\frac{\pi}{a},m=0,\frac{n}{2}$&$
\e^{\ri\tilde{k}\frac{a}{\tilde{q}}}$&$
\e^{\ri\tilde{m}\frac{2\pi}{n}}$&$\Pi^U$&$\ket{\tilde{k}\tilde{m}\Pi^U}$\\
${_{\tilde{k}}}E_{\tilde{m}}$&${k=0,\tilde{q}\frac{\pi}{a},\
m\in(0,\frac{n}{2})\atop
k\in(0,\tilde{q}\frac{\pi}{a}),\
m\in(-\frac{n}{2},\frac{n}{2}]}$&$K_2(\frac{\tilde{k}}{\tilde{q}})$&$
M_2(\tilde{m})$&$O_2$&${\ket{\tilde{k}\tilde{m}}\atop
\ket{-\tilde{k},-\tilde{m}}}$\\ \br
\end{tabular}\end{indented}
\end{table}

\begin{table}
\caption{Irreducible representations of $\bL_{\cZ\cA}$ classified by
the $km$ quantum numbers. For the irreducible representations denoted
in the first column allowed values of $k$ and $m$ are given in the
second column, and then the matrices of the generators follow.
Finally, SAB is given in terms of quantum numbers. Only for $n$ even
${_\pi}E^{\Pi^U}_{n/2}$ appears. The four dimensional matrices are
$M_4(m)=I_2\otimes M_2(m)$, $K_4(k,m)=K_2(k)\otimes M_2(m)$ and
$V_4=I_2\otimes O_2$.} \label{TCir13I}\begin{indented} \item[]
\footnotesize\rm
\begin{tabular}{@{}lllllll}\br
IR&$(k,m)$&$(C^1_{2n}|\frac{a}{2})$&$C_n$&$U$&$\sigma_x$&SAB\\ \mr
${_0}A/B^{\Pi^h}_m$&$k=0,m=0,n$&$\e^{\ri m\frac{\pi}{n}}$&$1$&$
   \Pi^h\Pi^v$&$\Pi^v$&$\ket{0m\Pi^h\Pi^v}$\\
${_0}E^{\Pi^h}_m$&$k=0,m\in(0,n)$&$M_2(\frac{m}{2})$&$M_2(m)$&$
\Pi^hO_2$&$O_2$&${\ket{0m\Pi^h}\atop \ket{0,-m,\Pi^h}}$\\
${_k}E^{A/B}_m$&${k\in(0,\frac{\pi}{a}),m=0,n\atop
k=\frac{\pi}{a},m=0}$&$
\e^{\ri\frac{m\pi}{n}}K_2(\frac{k}{2})$&$I_2$&$\Pi^vO_2$&$\Pi^vI_2$&$
{\ket{km\Pi^v}\atop \ket{-k,m,\Pi^v}}$\\
${_\pi}E^{\Pi^U}_\frac{n}{2}$&$k=\frac{\pi}{a},m=\frac{n}{2}$&$ \ri
K_2(\frac{\pi}{2a})$&$-I_2$&$\Pi^UI_2$&$O_2$&$
{\ket{\frac{\pi}{a},\frac{n}{2},\Pi^U}\atop
\ket{\frac{\pi}{a},-\frac{n}{2},\Pi^U}}$\\
${_k}G_m$&${k\in(0,\frac{\pi}{a}),m\in(0,n)\atop
k=\frac{\pi}{a},m\in(0,\frac{n}{2})}$&$K_4(\frac{k}{2},\frac{m}{2})$&$
M_4(m)$&$O_4$&$V_4$&$
{\ket{km},\ket{k,-m},\atop\ket{-k,m},\ket{-k,-m}}$\\ \br
\end{tabular}\end{indented}
\end{table}

Due to $U$ axis in $\bL_\cC$ and $\bL_{\cZ\cA}$ the linear quasi
momentum $k$ runs over the irreducible domain $[0,\pi/a]$ (and
$\tilde{k}\in[0,\tilde{\pi}]$ with $\tilde{\pi}=\tilde{q}\pi/a$)
being the half of Brillouine zone. In its interior the
representations are grouped in the $k$-series differing by $m$ (or
$\tilde{k}$-series differing by $\tilde{m}$). At the boundaries some
representations of $\bL_\cC$ have $U$-parity quantum number
$\Pi^U=\pm1$ denoted as the superscript $+/-$. As for the group
$\bL_{\cZ\cA}$ the superscript $+/-$ stands for $\sigma_h$-parity
(with respect to the horizontal mirror plane $\sigma_h=U\sigma_x$)
$\Pi^h=\pm1$, except for the representations ${_\pi}E^\pm_{n/2}$
where it corresponds to $\Pi^U=\pm1$; additionally, some of the
$k$-series and boundary representations of $\bL_{\cZ\cA}$ have
quantum number of $\sigma_x$-parity $\Pi^v=\pm1$, denoted by $A/B$,
respectively. The representations are given by the matrices of the
group generators. In the tables $I_n$ and $O_n$ stand for
$n$-dimensional diagonal and off-diagonal unit matrices respectively,
while the two dimensional diagonal matrices are
$M_2(m)=\mbox{diag}[\e^{\ri m2\pi/n},\e^{-\ri m2\pi/n}]$,
$K_2(k)=\mbox{diag}[\e^{\ri ka},\e^{-\ri ka}]$.

\end{document}